\documentclass{aastex}
\usepackage{psfrag}
\usepackage{graphicx,shapepar}

\newcommand{\kms}{\ensuremath{\mathrm{km\ s}^{-1}}}
\newcommand{\nii}{[\ion{N}{2}]}
\newcommand{\oiii}{[\ion{O}{3}]}
\newcommand{\heii}{\ion{He}{2}}
\newcommand{\vhel}{$V_{\textrm{\scriptsize hel}}$}
\newcommand{\vsys}{$V_{\textrm{\scriptsize sys}}$}
\newcommand{\grados}{$\,^\mathrm{o}\,$}
\title{The outflows and 3D structure of NGC 6337, a  planetary nebula with a close 
binary nucleus}

\author{Ma.\ T. Garc\'{\i}a-D\'{\i}az, D. M. Clark, J. A. L\'opez, W. Steffen \& M. G., Richer
  \affil{Instituto de Astronom\'ia, Universidad Nacional Aut\'onoma de
  M\'exico} Campus Ensenada, Ensenada, Baja California, 22800,
  M\'exico}
\affil{tere, dmclark, jal, wsteffen, richer@astrosen.unam.mx}

\begin{abstract}

NGC 6337 is a member of the rare group of planetary nebulae where a close binary nucleus has been  identified. The nebula's morphology and emission line profiles are both unusual, particularly the latter. We present a thorough mapping of spatially resolved, long-slit echelle spectra obtained over the nebula that allows a detailed characterization of its complex kinematics. This information, together with narrow band imagery is used to produce a three dimensional model of the nebula using the code SHAPE. The 3-D model yields a  slowly expanding toroid with large density fluctuations in its periphery that are observed as cometary knots. A system of bipolar expanding caps of low ionization are located outside the toroid. In addition, an extended high velocity and tenuous bipolar collimated outflow is found emerging from the core and sharply bending in opposite directions, a behavior that cannot be accounted for by pure magnetic launching and collimation unless the source of the outflow is precessing or rotating, as could be expected from a close binary nucleus.
\end{abstract}

\keywords{planetary nebulae: individual (NGC6337)  - ISM: jets and outflows - stars: binaries}

\begin{document}
\maketitle

\section{Introduction}
\label{sec:introduction}

To explain the complex morphologies and collimated outflows often observed in planetary nebulae the presence of toroidal magnetic fields and binary nuclei are commonly invoked \citep[e.g.,][]{ GS-Lo00, SoRa00}. However, from the observational standpoint  it has proven rather difficult to detect the firm supporting evidence for these theories. Recently, \citet{Ga-Daz08} have discussed the case of the planetary nebula NGC 1360 where evidence of a strong stellar magnetic field has been detected and magneto-hydrodynamical modeling has successfully reproduced its key kinematic and morphological features. In this work we now explore in detail the kinematic behavior  and morphological structure of NGC 6337, a nebula where convincing evidence of the presence of a close binary nucleus exists.
NGC 6337 (PN G349.3-01.1) appears as a thick ring with radial filaments and knots, a faint elliptical shell surrounds the ring or toroid and a conspicuous pair of condensations bright in \nii{} are located at P.A. $-$43\degr{} on opposite sides of the ring. The core of this nebula has been identified from time-resolved CCD (V-band) photometry \citep{Hillwig04} as a close binary nucleus with a period of 0.173 days. \citet{Hillwig04} assumes a primary mass of 0.6 M$_{\odot}$ to derive a mass for the companion star M$_2$$\leq 0.3$ M$_{\odot}$ and a binary separation a $\leq 1.26$ R$_{\odot}$, these characteristics indicate the possibility that the binary core underwent a common envelope phase. 

A  kinematic analysis of NGC 6337 has previously been made by \citet{Corradi00} via two long-slits across the nebula with position angles, P.A. = $-$39\degr{}
and P.A. = $-75$\degr.  From these data they interpreted the ring-like structure as
the waist of a pole-on bipolar PN. They also identified the expanding caps, which they labeled features A and B, located to the north-west (NW) and south-east (SE) and detected limited regions of high velocity that they tentatively identified as corresponding to a point-symmetric bipolar outflow. Here we extend that study by using 12 long-slit positions that fully sample the extent of the nebula and allow us to analyze in detail the strange structure of the line profiles and combine them with the morphological information to produce a morpho-kinematic 3-D model with SHAPE \citep{Steffen06} (http://www.astrosen.unam.mx/shape/) of the main nebula and its complex outflows that resolve the aspect-oriented complexity of this object.

The paper  is organized as
follows. In \S2 we present the observations, in \S3 we discuss the results,  \S4  describes the 3-D SHAPE model  of NGC 6337 and finally in \S5 we summarize the conclusions of this study.

%\newpage 
\section{Observations and Results} 
\label{sec:observations}

High-resolution spectroscopic observations and monochromatic images of NGC 6337 were
obtained at the Observatorio Astron\'omico Nacional at San Pedro
M\'artir, (SPM), M\'exico, on two observing runs 2006, July 20
and 2007 June 20 -- 23. These observations were taken using the
Manchester Echelle Spectrometer (MES-SPM) \citep{Meaburn03} on the 2.1
m telescope in a $f$/7.5 configuration.  This instrument was equipped
with a SITE-3 CCD detector with 1024 $\times$ 1024 square pixels, each
24 $\mu$m on a side.  We used a 90 \AA{} bandwidth filter to isolate
the 87th order containing the H$\alpha$ and \nii{} nebular emission
lines.  Two times binning was employed in both the spatial and
spectral directions. Consequently, 512 increments, each 0\farcs624{}
long gave a projected slit length of 5\farcm32 on the sky. We used
a slit of 150 $\mu$m{} wide ($\equiv$ 11 \kms{} and 1\farcs9) oriented
to a P.A. of $-$43\degr{} for the majority of the observations and a
P.A. $=$ 0\degr{} for one pointing across the center of the
nebula. All spectra and images were acquired using exposure times of
1800 s. The spectra were calibrated in wavelength against the spectrum of a Th/Ar arc lamp to an accuracy of $\pm$1 \kms{} when converted to radial velocity.
Deep images of the field  were also obtained with MES in its imaging configuration in
three different filters: \nii{} $\lambda$ 6584~\AA, H$\alpha$$+$\nii{}, and \oiii{} $\lambda$ 5007~\AA ~with bandwidths 
of 10~\AA, 90~\AA, ~and 50~\AA, respectively.

We reduced the data using standard IRAF\footnote{IRAF is distributed
by the National Optical Astronomy Observatories, which is operated by
the Association of Universities for Research in Astronomy, Inc. under
cooperative agreement with the National Science foundation} tasks to
correct bias, remove cosmic rays and calibrate the two-dimensional
spectra based on the comparison lamp spectra. All spectra presented in
this paper are corrected to heliocentric velocity (\vhel).

\begin{figure*}[!t]
\begin{center} 
  \includegraphics[width=1\textwidth]{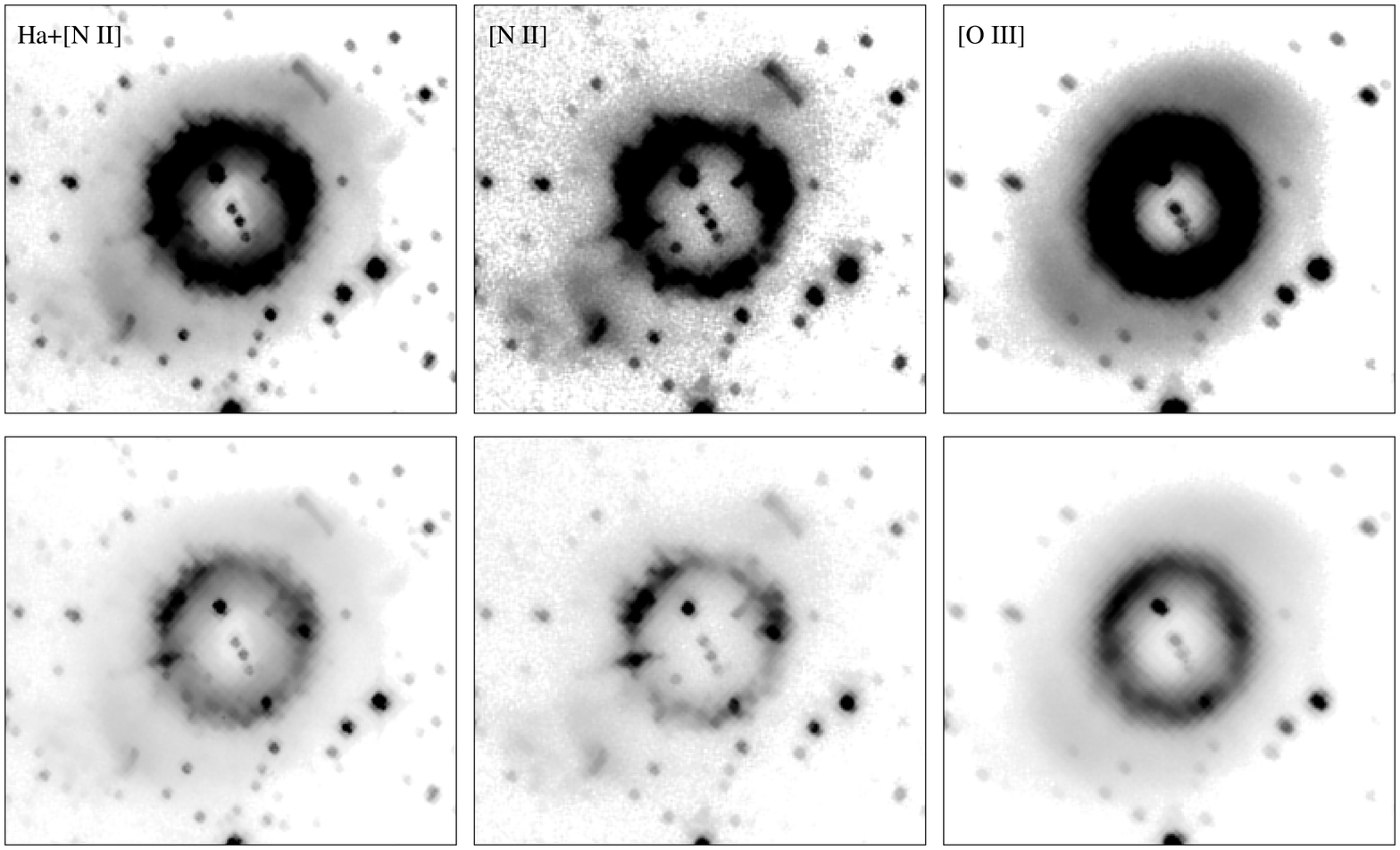} 
\caption{Images of NGC 6337 obtained with MES -- SPM in the light of  H$\alpha$$+$\nii{} (left column), \nii{} $\lambda$ 6584~\AA ~(center) and \oiii{} $\lambda$ 5007~\AA ~(right), these are displayed at two different dynamic ranges to show the rich structure of the nebula}
\label{fig:images}
\end{center}
\end{figure*}

\begin{figure*}[!t]
\begin{center} 
  \includegraphics[width=1\textwidth]{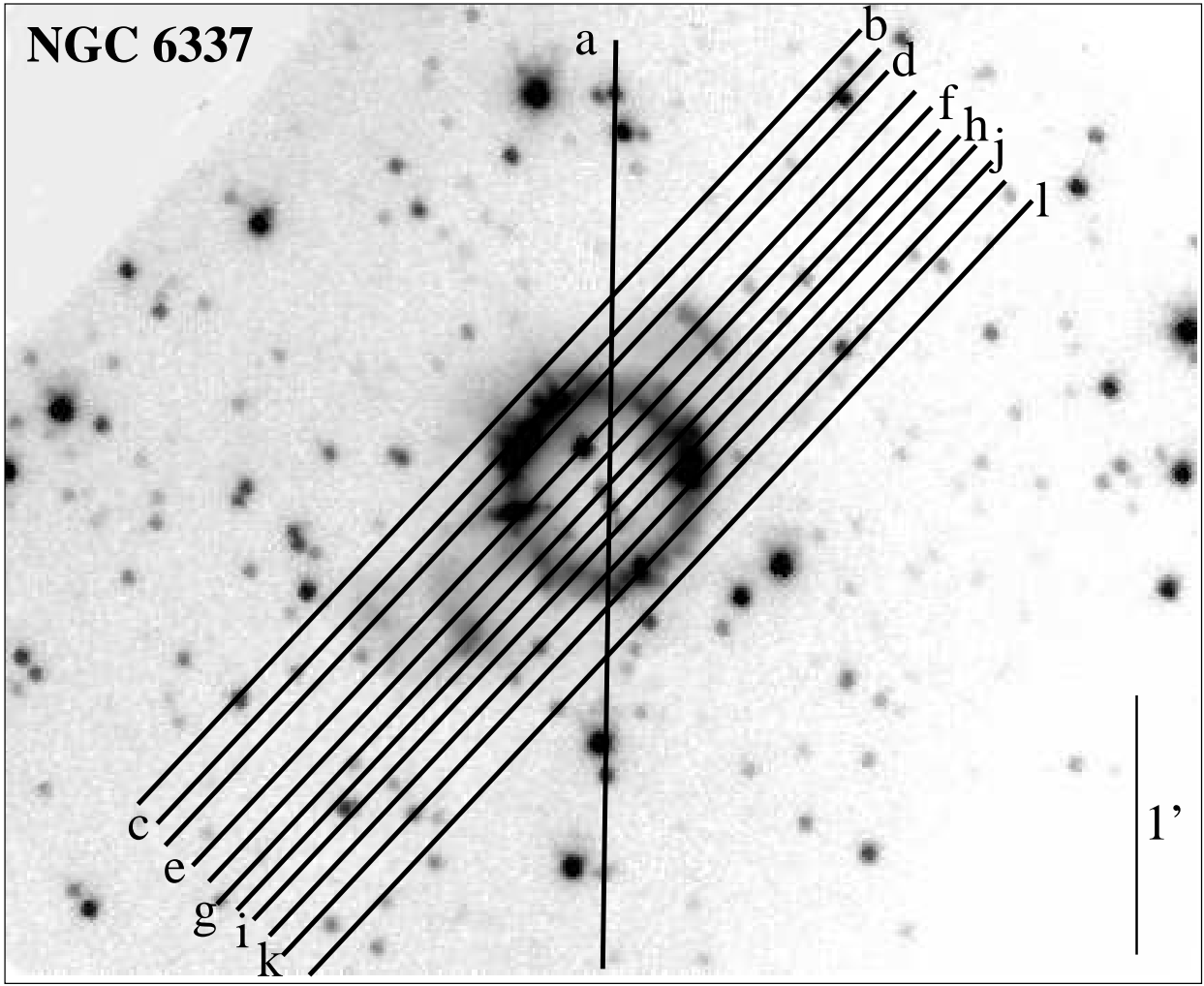}
  \caption{The location of each slit position is indicated and labeled on the
    \nii{}  image of NGC 6337}
  \label{fig:imageslit}
\end{center}
\end{figure*}

\begin{figure*}[!t]
\begin{center} 
  \includegraphics[width=1.05\textwidth]{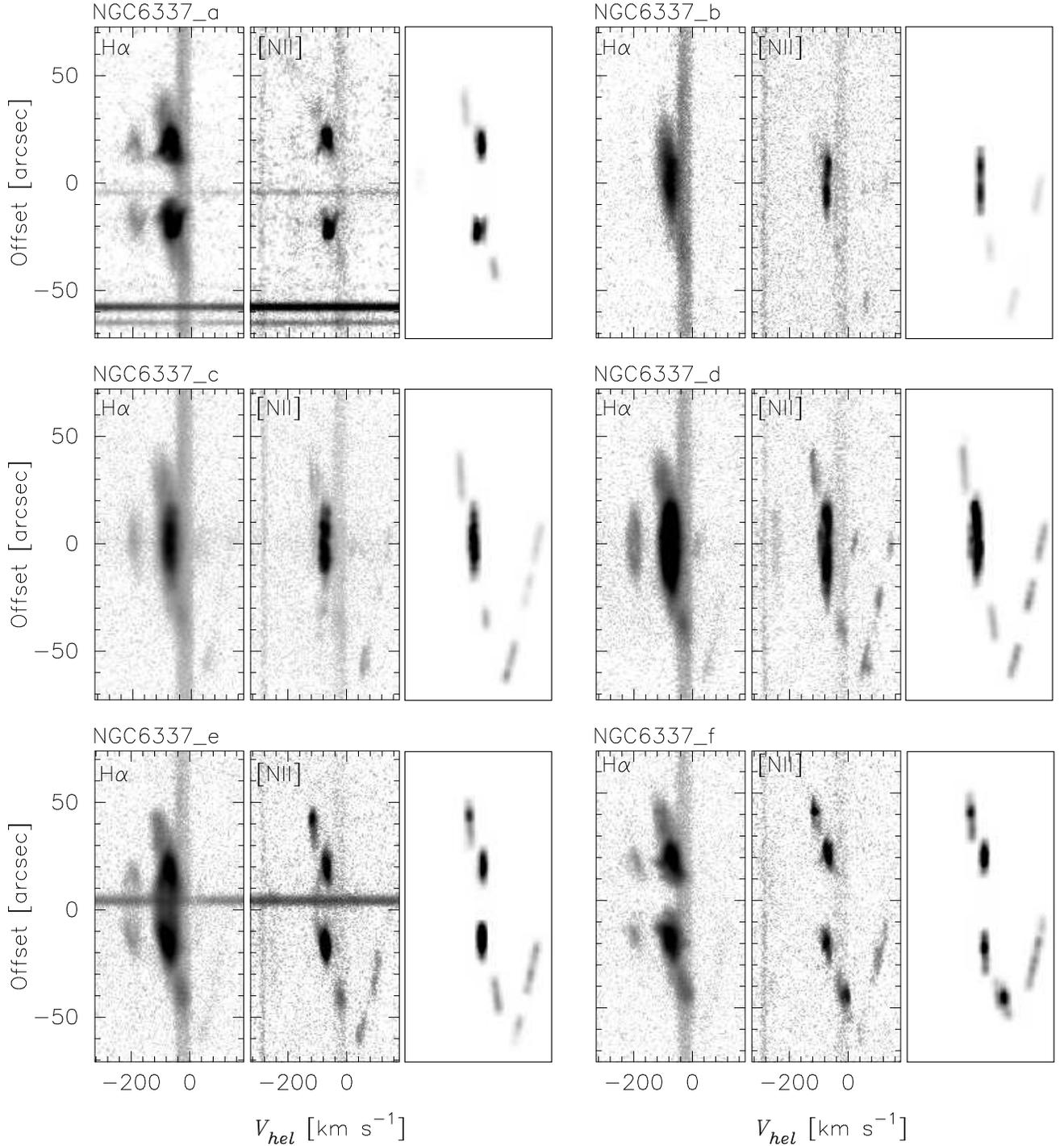}
  \caption{Mosaic of bi-dimensional position -- velocity (P -- V) arrays. For each slit position there is a group of three P -- V arrays, the observed H$\alpha$ and \nii~$\lambda$~6583~\AA{} line profiles and the corresponding synthetic  \nii~$6583$~\AA{}  P -- V array modeled with SHAPE.  Slit positions a -- f are shown here. }
    \label{fig:spectra_a}
\end{center}
\end{figure*}

\begin{figure*}[!t]
\begin{center} 
  \includegraphics[width=1.05\textwidth]{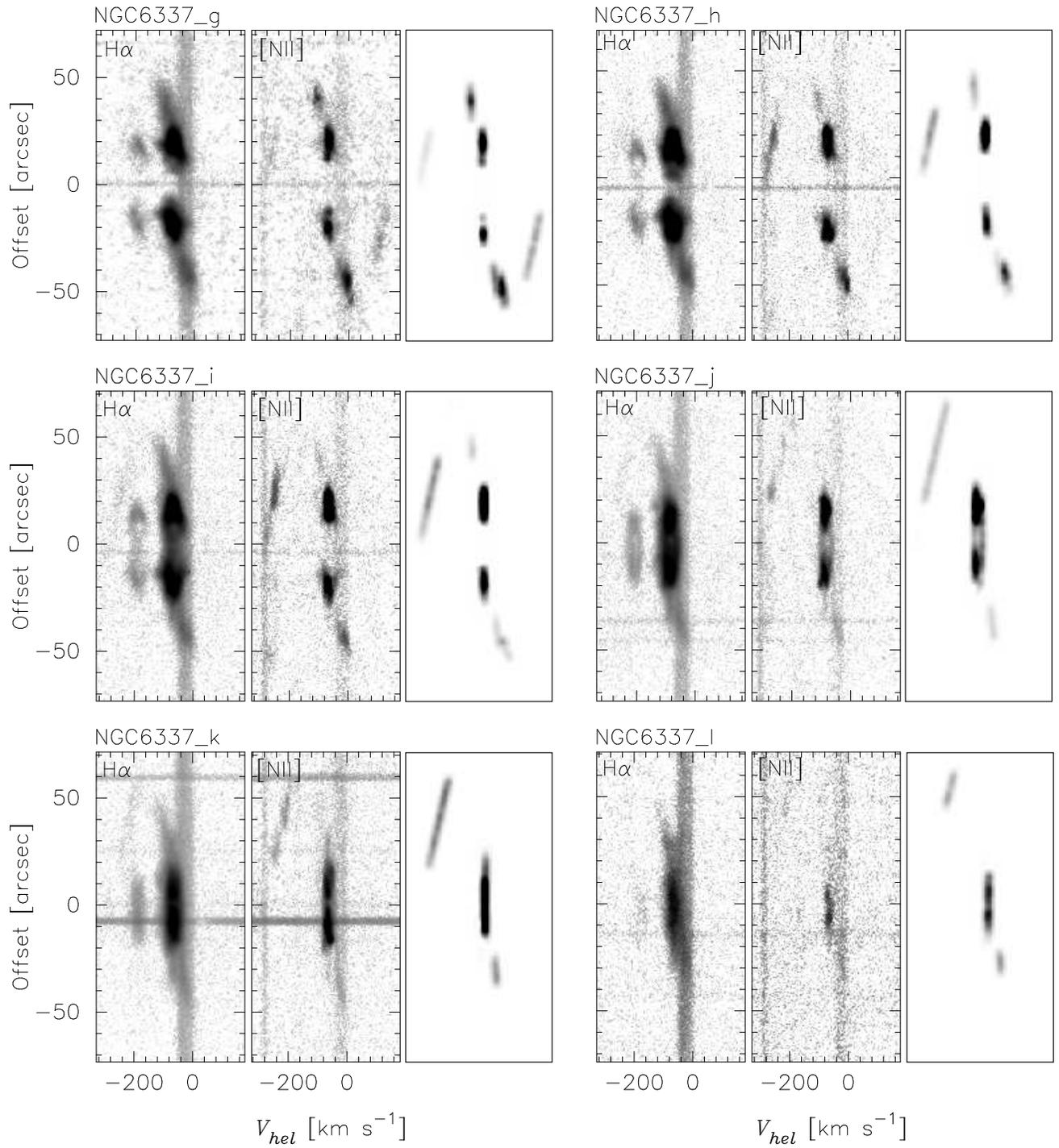}
  \caption{As in Figure 3, but for slit positions g -- l.}
  \label{fig:spectra_b}
\end{center}
\end{figure*}

Figure~\ref{fig:images} is a mosaic of  images of NGC 6337, namely H$\alpha$$+$\nii{} , \nii{} $\lambda$ 6584~\AA, ~and \oiii{} $\lambda$ 5007~\AA. Since the original images are deep, with high signal to noise ratios, here they are displayed at two different digital contrast levels to show the rich structure of this nebula. In Figure~\ref{fig:imageslit}, the slit positions are indicated and labeled on the
\nii{} image of NGC 6337 taken at SPM. We obtained 11 consecutive positions 
with a P.A. = $-$43\degr{}  across the nebula and one position at  P.A. = 0\degr{} to 
encompass all its main components. The bi-dimensional emission line spectra or position -- velocity 
(P -- V) arrays from slit positions a -- f are shown in Figure~\ref{fig:spectra_a}, and from g -- l in Figure~\ref{fig:spectra_b}.

\section{Discussion} 
\label{sec:discussion}

The images in Figure~\ref{fig:images} show the bright main ring, its filamentary and knotty structure and the bright, low ionization filaments  located to the northwest and southeast of the ring and that are referred to here as caps. In addition, the H$\alpha$$+$\nii{}  image shows a faint elliptical outer shell which is also observed in the \oiii ~$\lambda$ 5007~\AA ~image where it displays a brightness distribution that resembles a slight "S" shape, sometimes associated with bipolar envelopes. A tenuous, high-speed, bipolar outflow that is clearly present in the long-slit spectra (see below) cannot be identified from these images and is revealed by the SHAPE modeling of the nebula to be associated with some of the diffuse material that surrounds the caps, most clearly seen in the \nii{} $\lambda$ 6584~\AA ~image.

In Figures 3 and 4 a mosaic of three individual bi-dimensional arrays are presented for each slit position. The observed H$\alpha$ and \nii{}  line profiles are on the left and to the right are the  \nii{}  synthetic line profile produced by the SHAPE model.
The observed spectra are shown here corrected for heliocetric
velocity and the offset in declination is with respect to the central
star.  The heliocentric systemic velocity, \vsys, is
$-$70.56 \kms{} calculated using slit g, which passes through the
central star, this velocity is in good agreement with the velocity
derived by \citet{Corradi00} and \citet{Meath88}.

The spectra show clearly the signature of a thick ring, or more properly a toroid, in the form of symmetric bright knots of emission that appear where the slit crosses it.  The toroid shows line profiles with a wedge-like shape indicating expansion of the section of the toroid facing the star, where line splitting is observed. For slit a, the average splitting in the \nii{} line emission  from the top and bottom sides of the toroid amounts to 39.3 \kms{} or 19.7 \kms{} of expansion velocity along the line of sight. There is some tenuous material close to the inner borders of the toroid that share this expansion, as revealed by the velocity ellipses in the \nii{} line emission from slits e and j that join the emission knots from the toroid. 

The toroid shows a very slow projected radial expansion, in the order of only 1 - 2 \kms{} on average. This very slow radial expansion together with the nearly perfect circular shape of the toroid indicates that it has a very small tilt with respect to the plane of the sky ($\leq 10$\grados) and that we are looking at it nearly face on.  A crude estimate of the deprojected expansion velocity for the ring yields about 11.5 \kms{}, assuming a 10\grados tilt with respect to the plane of the sky. The angular outer radius of the ring is 24\arcsec, adopting a distance D = 1.3 pc  \citep{Sta93} to  NGC 6337, its linear radius becomes 0.15 pc  $\equiv 4.66 \times 10^{17}$ cm. These values yield a kinematic age for the ring of $1.2 \times 10^{4}$ years. 

The toroid and its inner region contain high excitation gas, revealed by the presence of the \heii{}  $\lambda$ 6560 line emission. Positions e, h, and i show a bipolar type structure in the line emission of this ion. It is however curious that the regions closest to the star, see slits a, f, and g, do not show projected \heii{}  emission, as if a cylindrical cavity perpendicular to the plane of the toroid is indeed fairly void of material close to the core. A cavity produced by an isotropic  free flowing wind region would be expected to contain the high excitation gas at its edge and observable in projection over the star, which is not the case.  The presence of this cavity is perplexing and to discern its origin  will require additional information and modeling (out of the scope of the present work).
 
The caps are readily recognized in the spectra as the knotty extensions located immediately outside the ring emission regions, prominent in slit positions e to i.  They are reminiscent of FLIERS \citep{Balick93} though in this case they seem to be possibly related to mass ejections associated with the fast bipolar outflows. The cap to the northwest is blueshifted and slit f that cuts across the center of this filament yields a heliocentric  velocity \vhel =  $-$44.9 \kms{} with respect to \vsys. The corresponding velocities along the filament and just outside of it on both sides (slits d to h) show a decreasing trend in velocity from the northeast end of the cap to its  southwest tip with values of \vhel = $-$47.0 \kms{} (slit d), $-$45.1 \kms{} (slit e), $-$40.6 \kms{} (slit g), and $-$31.0 \kms{} (slit h) with respect to \vsys. This effect might indicate that the filament that forms the northwest cap (and associated surrounding material) is twisted or tilted with respect to the sightline. The cap on the southeast is redshifted with respect to the systemic velocity and slit g yields \vhel = $+$64.7 \kms{} for this region. The slits on either side of slit g, slits f and h, provide very similar values, \vhel  = $+$64.9 \kms{}  and $+$64.8 \kms{} respectively with respect to \vsys. Considering the inclination angles with respect to the line of sight derived from the SHAPE model  for the caps, their kinematic ages range from $ 1 - 2 \times 10^3$ years, i.e. these are relatively young structures compared to the ring. The diffuse envelope material that surrounds the caps is apparent in the H$\alpha$ line profiles of Figures 3 and 4 where it is seen to follow the same velocity trend as the caps.

An outstanding feature of the \nii{} emission line profiles in NGC 6337 are the very high velocity components that run from \vhel = $-$210 \kms{}  to \vhel = + 215 \kms{} with respect to the systemic velocity, \vsys, corresponding to bipolar, collimated, jet-like outflows. Slits b to g show the line profiles with the redshifted components (east, southeast side of toroid) spanning a range of  $-$275 \kms{} $\leq{}$ \vhel{} $\leq{}-$ 214 \kms{}   whereas slits g to l show those blueshifted (west, northwest side of toroid)
within a range of +40 \kms{} $\leq{}$ \vhel{} $\leq{}$ +145 \kms{}; the line profile from slit g shows both, i.e. slit g, that passes close to the central star shows both sides of the high-speed bipolar outflow, covering a velocity range of over 400 \kms{} . The high-velocity components in all slits display their maximum velocity close to the star and this decreases with distance from it, contrary to what is usually observed in the so-called Hubble flows where the outflow increases its velocity with distance from the source \citep[e.g.,][]{Meaburn08} .

\section{SHAPE modeling}
\label{sec:shape}

In order to disentangle the 3-D geometry and kinematic structure of NGC 6337, we used the program SHAPE  \citep{Steffen06}. SHAPE is a morpho-kinematic modeling tool that allows the user to reconstruct the 3D structure and observed spectral line profiles using expanding geometrical forms. SHAPE uses as reference monochromatic images and observed position -- velocity diagrams to reproduce the 3-D structure and kinematics of the object.  Particles are distributed over a surface, or throughout a specified
volume, and are assigned a specific velocity law and relative brightness. Several particle systems can be used and each can be assigned different velocity laws to form a complex object. The resultant 2-D image and spectral information are then rendered from the 3-D model and compared with the real data. 

\begin{figure*}[!t]
\begin{center} 
  \includegraphics[width=1.05\textwidth]{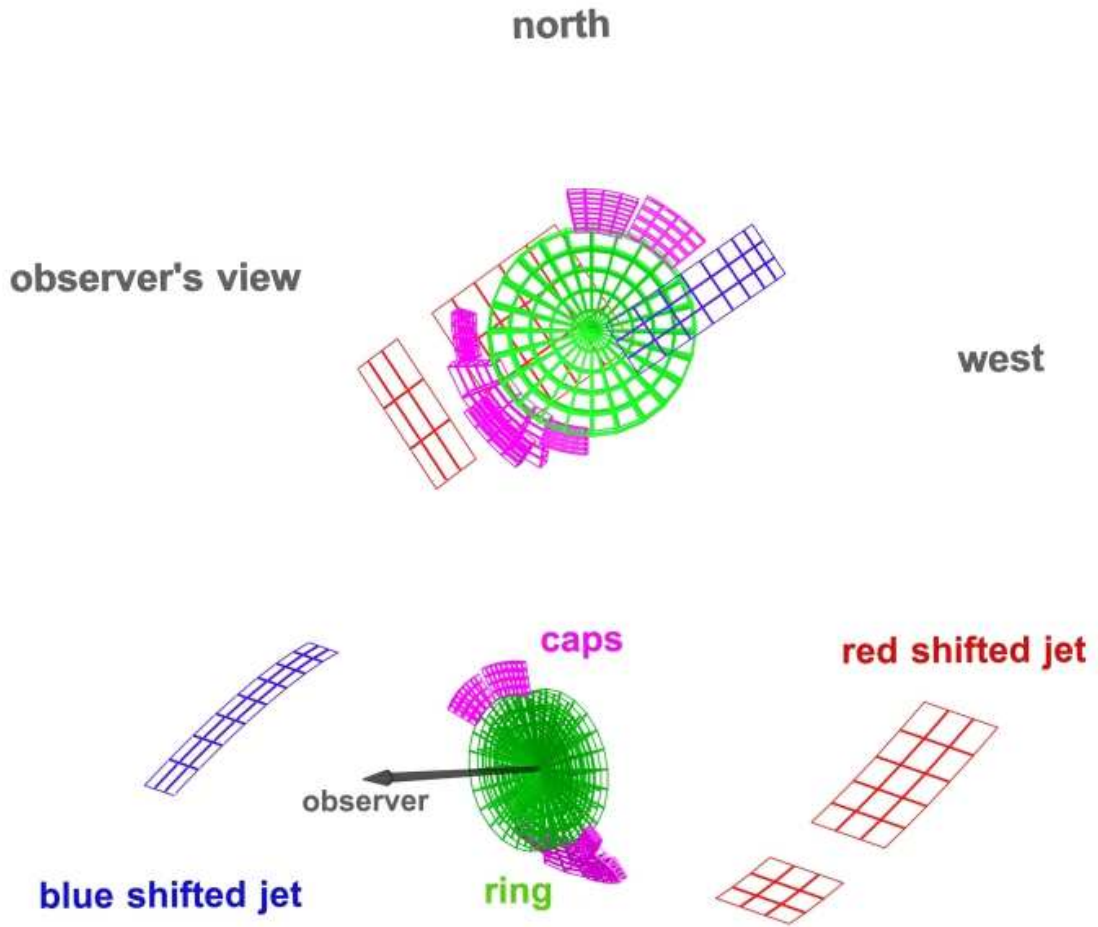}
  \caption{SHAPE mesh model of NGC 6337 before rendering, shown at two different orientations}
  \label{fig:shape_raw}
\end{center}
\end{figure*}

To model NGC 6337, we used the \nii{} image and position-velocity spectra since all the main kinematic components are present here. Our model was built with a torus and conic surfaces for its knotty  structure. Segments of spheres were used to model the caps. The bipolar collimated outflows were built from elongated cylinders, flattened and bent. The observations indicate that the projected width of the redshifted section of the bipolar outflow is slightly wider than its blue counterpart and it has been modeled accordingly. This difference in widths may result from projection effects along the line of sight such as a slightly twisted bipolar outflow, as may be indicated by the kinematic analysis of the northwest cap (see section 3) however, since a projected twisting effect of the outflows cannot be disentangled from the emission line profiles of the collimated outflows, this potential twisting has not been considered in the model. Figure~\ref{fig:shape_raw} shows the resultant 3-D mesh of the model before rendering.

\begin{figure*}[!t]
\begin{center} 
  \includegraphics[width=0.5\textwidth]{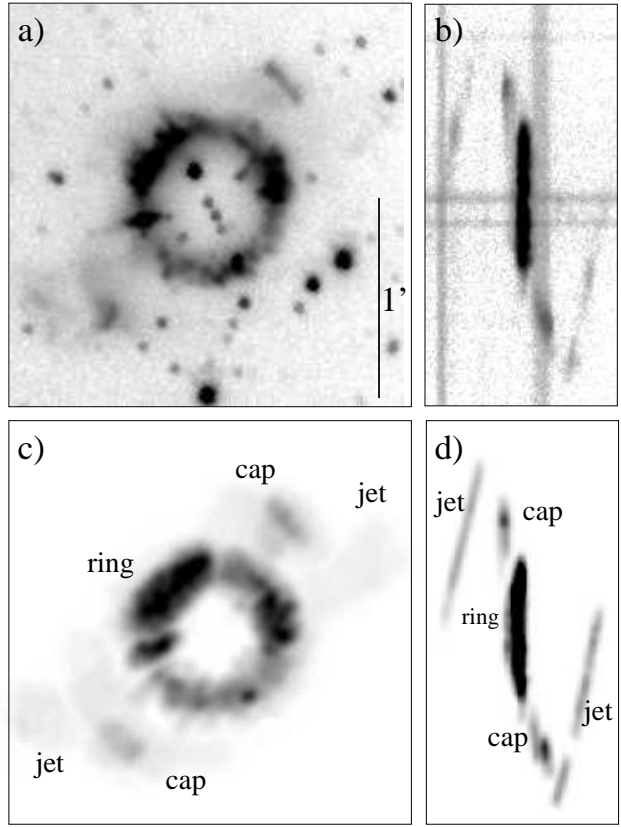}
  \caption{Top panels, frames a and b, show the \nii{} image of NGC 6337 and  the integrated, observed
   \nii{}  P - V arrray from slits b -- l. Lower panels, frames c and d, show the synthetic image and integrated synthetic  \nii{}  P - V array from slits b -- l, modeled with SHAPE. The main components are labeled in both the image and the integrated line profile.}
  \label{fig:model}
\end{center}
\end{figure*}

\begin{figure*}[!t]
\begin{center} 
  \includegraphics[scale=0.9]{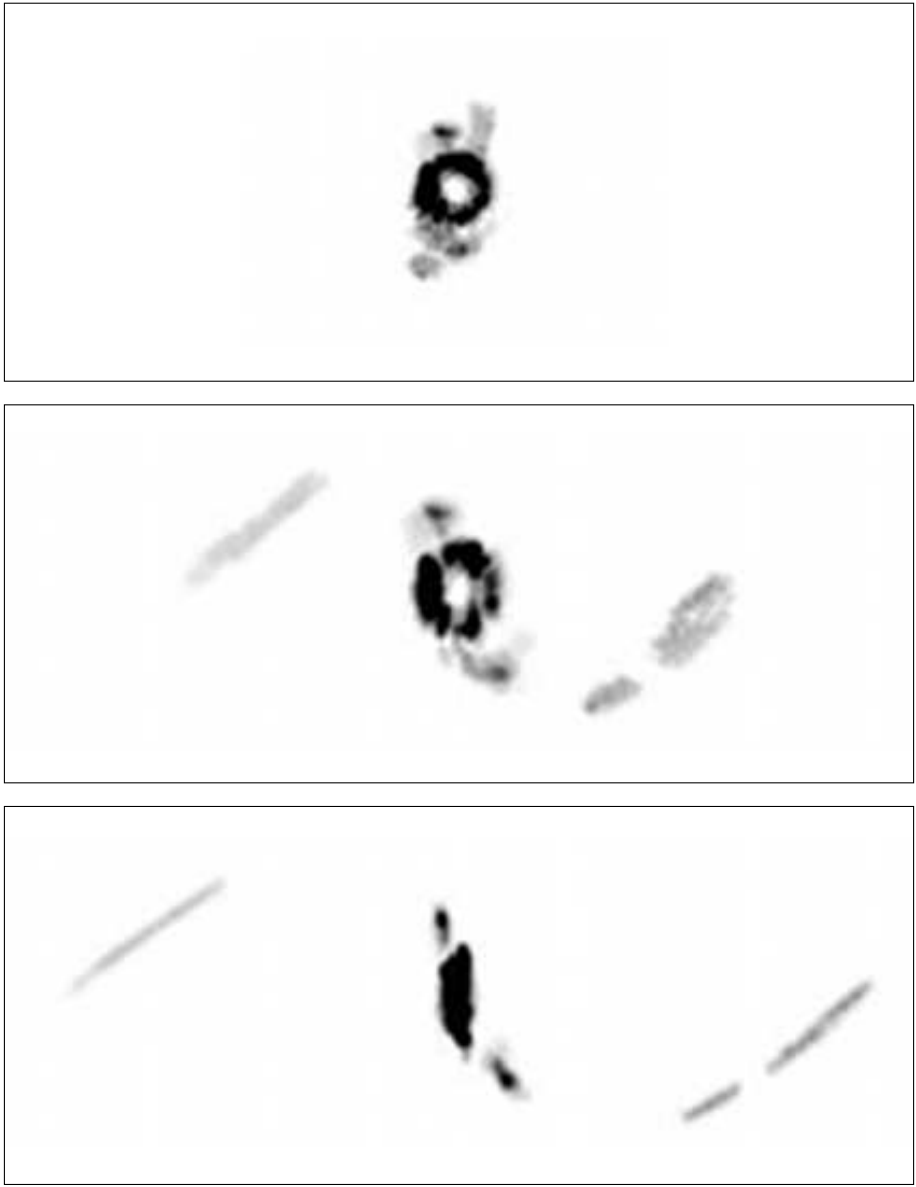}
  \caption{3-D rendered representation of NGC 6337 displayed at various viewing angles. The top frame is  shown with north rotated 43\degr{} counter-clockwise, equivalent to having slits b -- l aligned vertically. The next two panels are rotated on the y-axis (defined along the vertical direction) by 45\degr{} (center) and 90\degr (bottom) clockwise, respectively. }
  \label{fig:rendering}
\end{center}
\end{figure*}

The results of the final rendered model are shown in Figure~\ref{fig:model} where they are compared to the integrated observed data.  Panel a shows the \nii{} SPM image of NGC 6337 and panel b shows the sum of all of the observed spectral line profiles for slits b -- l, simulating a slit equivalent to the width of the region sampled by slits b -- l, but preserving the spectral resolution. Panels c and d show the corresponding  rendered image and integrated synthetic P -- V array  from the same slits b -- l. The main  structural components are labeled in the model image and integrated line profile. The individual rendered synthetic \nii{} P -- V arrays from the model are shown in Figures 3 and 4 next to the observed ones, where it is apparent that they provide a reasonable match in all instances. Finally, we present the 3-D rendered representation of NGC 6337 in Figure~\ref{fig:rendering} 
displayed at various viewing angles. The first frame is shown with the north rotated 43\degr{} counter-clockwise, equivalent to having slits b -- l aligned vertically. The next two panels are
rotated on the y-axis by 45\degr{} and 90\degr{}, clockwise, respectively. 
As an additional test of the overall goodness of our model we have produced synthetic line profiles from slits located at  P.A. = $-39$\degr and P.A. = $-75$\degr, corresponding to those observed by \citet{Corradi00}, obtaining also a good match with their observations.

 The present SHAPE model is able to reproduce the basic 2-D morphology and the set of emission line profiles that provide a representation of the third dimension of the nebula through the radial velocity component. Slightly different geometric forms could have been used to build up the final model, but in the end there is only a very limited set of solutions that are able to replicate the complex P -- V diagrams of this nebula. A key advantage of SHAPE for objects like NGC 6337 is its ability to model independent structures, each with its own velocity law, and then merge them into a single product. We have used velocity laws of the type $v = k \cdot r/r_o$ where k is a constant, $r$ is the distance from the source and r$_0$ is the distance at which the velocity $k$ is reached. The values for $k$ and $r_0$ have been chosen to match the observed velocities for the various components of NGC 6337 and to provide reasonable distance scales along the line of sight. However, the model cannot place restrictions on, for example, the length of the bipolar collimated outflows or its detailed structure nor on the thickness of the toroid or the precise distance of the caps from the toroid. Nevertheless, the present model allows a good understanding of the complex structure and outflows of this object that otherwise are rather difficult to visualize.

\section{Conclusion} 
\label{sec:conclusion}

A thorough analysis of the kinematic structure and morphology of the planetary nebula NGC 6337 has been carried out. The nebula is composed of a conspicuous thick ring or torus and fast (\vhel{} $\gtrsim$ 200 \kms{} ) bipolar, collimated, outflows that bend in opposite directions in a point-symmetric way. The ring does not seem to be  the collimating agent for the bipolar outflows. The torus is slowly expanding radially, at a rate of only a few kilometers per second, and also shows an internal expansion of $\approx 20$ \kms{}. It should be noted that thick rings or toroids are uncommon as main morphologies of planetary nebulae, although equatorial density enhancements are common in axi-symmetric nebulae such as ellipsoidal and bipolar. The faint outer envelope of the nebula seems to follow the emergence of the poloidal outflows silhouetted in projection and it is likely that this envelope material  is what remains of incipient  bipolar lobes blown at an early stage of  development of the nebula. In this case only traces of these lobes, close to the ring, remain now since no additional indication of their presence is apparent in the narrow band images nor in the emission line spectra. The caps may be mass concentrations of this material that have been pushed aside by the collimated outflows in their way out or a late bipolar episodic event, given their shorter age.

Considering the current parameters derived by \citet{Hillwig04} for the binary core, it is likely that this underwent a common envelope episode that may have influenced the formation of the equatorial density enhancement into a thick ring. The region close to the core appears as a cylindrical cavity void of material; it is unclear how this cavity may have formed. The rich, knotty structure apparent in the thick ring has most likely resulted as a consequence of instabilities produced by  the erosive interaction with the radiation field and winds from the core throughout the evolution of the system.  We confirm the suggestion by \citet{Corradi00} of the presence of a point-symmetric, bipolar, collimated outflow;  SHAPE modeling of our data has revealed the structure of such a jet for the first time.  Although the binary interaction may have spun up the white dwarf central star, favoring a magnetic launching and collimation of the outflows, it is unlikely that a magneto-hydrodynamic mechanism can be solely responsible of the extreme bending of the bipolar, point symmetric, collimated outflows.  \citet{GS-Lo00} have shown in these cases that although point-symmetric structures can be obtained from MHD models,  the collimated outflows interact with the inner walls of the wind blown lobes to form point-symmetric structures, as in the case of the planetary nebula Hb 5. The extreme characteristics of the bipolar, collimated outflows in NGC 6337 require them to emerge from a rotating or precessing source, provided here in a natural way by the close binary core, in addition to a MHD collimating agent.  With all these characteristics, NGC 6337 can be considered an archetypical example of the potential influence of a close binary core on the evolution of a planetary nebula. 

\acknowledgments

MTG-D and DMC gratefully acknowledege the support of  postdoctoral grants from UNAM.
This research has benefited from the financial support of DGAPA-UNAM through grants
IN116908, IN108506 \& IN108406 and CONACYT grant 49447. We acknowledge the excellent support of the technical personnel 
at the OAN-SPM, particularly Gustavo Melgoza, Felipe Montalvo and Salvador Monroy, who 
were the telescope operators during our observing runs.

\bibliographystyle{astroads}

\end{document}